\documentclass[12pt]{iopart}
\usepackage{graphics}
\usepackage{graphicx}
\input{epsf}
\def\l{\Lambda}
\def\rhoc{\rho_{0{\rm c}}}

\def\etal{et al.}

\def\eg{e.g.}

\def \lleq {\lower0.9ex\hbox{ $\buildrel < \over \sim$} ~}
\def \ggeq {\lower0.9ex\hbox{ $\buildrel > \over \sim$} ~}

\def\omt{\Omega_{0 \rm m}}
\def\om{\Omega_{\rm m}}
\def\half{{1\over 2}}

\def\spose#1{\hbox to 0pt{#1\hss}}
\def\simle{\mathrel{\spose{\lower 3pt\hbox{$\mathchar"218$}}
     \raise 2.0pt\hbox{$\mathchar"13C$}}}
\def\simge{\mathrel{\spose{\lower 3pt\hbox{$\mathchar"218$}}
     \raise 2.0pt\hbox{$\mathchar"13E$}}}

\def\apj{Astroph.~J.~}
\def\mn{Mon.~Not.~Roy.~Ast.~Soc.~}
\def\asta{Astron.~Astrophys.~}
\def\aj{Astron.~J.~}
\def\prl{Phys.~Rev.~Lett.~}
\def\prd{Phys.~Rev.~D~}

\def\plb{Phys.~Lett.~B~}
\def\jetpl{JETP ~Lett.~}

\def\beq{\begin{equation}}
\def\eeq{\end{equation}}
\def\ber{\begin{eqnarray}}
\def\eer{\end{eqnarray}}

\newcommand{\sq}{\lower.25ex\hbox{\large$\Box$}}

\begin{document}

\title{The case for dynamical dark energy revisited}
\author{
Ujjaini Alam${}^{a}$, Varun Sahni${}^{a}$
and A. A. Starobinsky${}^{b}$}

\address{${}^{a}$ Inter-University Centre for Astronomy \& Astrophysics,
Pun\'e 411 007, India}
\address{${}^{b}$ Landau Institute for Theoretical Physics,
119334 Moscow, Russia}
\date{\today}

\begin{abstract}
We investigate the behaviour of dark energy using the recently
released supernova data of Riess \etal, 2004 and a model independent
parameterization for dark energy (DE).  We find that, if no priors are
imposed on $\omt$ and $h$, DE which evolves with time provides a
better fit to the SNe data than $\l$CDM. This is also true if we
include results from the WMAP CMB data. From a joint analysis of
SNe+CMB, the best-fit DE model has $w_0 \lleq -1$ at the present epoch and 
the transition from deceleration to acceleration occurs at
$z_T = 0.39 \pm 0.03$. However, DE evolution becomes weaker if the
$\l$CDM based CMB results $\omt = 0.27 \pm 0.04$, $h = 0.71 \pm 0.06$
are incorporated in the analysis. In this case, $z_T = 0.57 \pm 0.07$.
Our results also show that the extent of DE evolution is sensitive to
the manner in which the supernova data is sampled.
\end{abstract}

% for PACS codes, see http://publish.aps.org/PACS/pacs99.html
%\pacs{PACS numbers:
%  98.80.Es, %Observational cosmology
%  98.80.Cq, %Particle-theory and field-theory models of the early Universe
%  98.80.Hw} %Mathematical and relativistic aspects of cosmology;
% 97.60.Bw} %Supernovae
\bigskip

\section{Introduction}

Supernova observations \cite{SN1,SN2} were the first to suggest that
our universe is currently accelerating.  Subsequently, a combination
of results from cosmic microwave background (CMB) experiments and
observations of galaxy clustering served to strengthen this world view
\cite{spergel03,tegmark03}, and it is now believed that as much as 2/3
of the total density of the universe is in a form which has large
negative pressure and which is usually referred to as dark energy
(DE).

The earliest theoretical model of DE -- the cosmological constant
($\l$) -- satisfied $w \equiv p/\rho = -1$.  Since the energy density
in $\l$ does not evolve, its present value $\rho_\l \simeq 10^{-47}$
GeV$^4$ is also its initial value. As a result, the ratio
$\rho_\l/\rho_r$, where $\rho_r$ is the radiation density, had the
miniscule value $10^{-123}$ at the Planck time. The enormous amount of
fine tuning this might involve led theorists to suggest that, like
other forms of matter in the universe, DE density too may show
significant time-evolution. However, this argument for significant
variability of $\l$ with redshift is questionable. Actually, it is a
variant of the Dirac's large number hypothesis that proved not to be
valid in our Universe. For example, the density of water has also a
miniscule value of $10^{-93}$ in Planck units (though, of course,
larger than that of $\rho_{\l}$), and we know that its relative change
from present time up to redshifts of the order of one (due to possible
variations of the fine structure constant and the electron and proton
masses) is less than $10^{-5}$. So, based on the argument above, it
would be wrong to assume that dark energy density should significantly
change with redshift.

Another, more reliable reason to suggest a time dependent form of DE
lies in the fact that our current accelerating epoch is unlikely to
have been unique.  In fact there is considerable evidence to suggest
that the universe underwent an early inflationary epoch during which
its expansion rapidly accelerated under the influence of an `inflaton'
field which, over sufficiently small time scales, had properties
similar to those of a cosmological constant.  Inspired by inflationary
cosmology, quintessence models invoke a minimally coupled scalar field
to construct a dynamically evolving model for DE.

Recent years have seen a flurry of activity in this area and there are
currently, apart from quintessence, at least a dozen well motivated
models of an accelerating universe in which dark energy is a
dynamically evolving quantity (see reviews
\cite{sahni02,ss00,sahni04}).  A simple categorization of DE models
could be as follows:

\begin{enumerate}

\item The cosmological constant, $w = -1$.

\item DE with $w = {\rm constant} \neq -1$
(cosmic strings ($w = -1/3$), domain walls 
($w = -2/3$). Quintessence with a sine hyperbolic 
potential \cite{ss00}).

\item Dynamical DE, $w \neq {\rm constant}$. (Quintessence, Chaplygin 
gas \cite{chap1}, k-essence \cite{mukhanov}, braneworld models 
\cite{DDG,ss02a,as02}, etc.)

\item DE with $w < -1$.  (scalar-tensor gravity models \cite{beps}, 
phantom models, braneworld cosmology etc. 
\cite{caldwell,innes,caldwell03,carroll03,melchiorri02}.)
\footnote{Note that the kinetic energy of the phantom scalar field is
negative, thus phantom models are strongly unstable with respect
to spontaneous creation of 4 particles: a particle-antiparticle pair
of this scalar field and a particle-antiparticle pair of any usual
matter quantum field (see \cite{cline03} for the most recent
investigation of this process). Scalar-tensor models of dark energy
with the effective Brans-Dicke parameter $\omega > -3/2$ are free from
this difficulty. However, a possible decrease of $w$ below $-1$ in
these models is typically of the order of $1/\omega$, and is therefore
very small, as follows from Solar System tests of general relativity.
In this context it is interesting that the Braneworld models suggested
in \cite{ss02a} can give rise to an effective equation of state for dark
energy which can be substantially less than $-1$, and yet not be plagued with
any instabilities at the classical or quantum level.}

\end{enumerate}

In view of the large number of possibilities for DE it would not be
without advantage to analyse the properties of DE in a model
independent manner. Such an approach was adopted by Alam \etal
\cite{asss03b} (henceforth Paper I) in which the Supernova data
published by Tonry \etal \cite{tonry03} and Barris \etal
\cite{barris03} was analysed using a versatile ansatz for the Hubble
parameter.  Paper I discovered that dynamical DE fit the SNe
observations better than $\l$CDM and these results found support in
the subsequent analysis of other teams
\cite{wangm,leandros,gong,daly}.  (Paper I referred to DE evolution as
`metamorphosis' since the DE equation of state appeared to
metamorphose from a negative present value $w_0 \lleq -1$ to $w \simeq
0$ at $z \simeq 1$.)

Recently Riess \etal  \cite{riess04} have reanalysed  some earlier SNe
data and also  published new  data relating  to 16  type Ia supernovae
discovered using  the Hubble  Space  Telescope (HST).  In the  present
paper we shall reconstruct the properties of DE using the new SNe data
set  (`Gold' in  \cite{riess04}). We shall   also use some  of the CMB
results obtained by WMAP in the later part of our analysis.

\section{Reconstructing Dark Energy}

Perhaps the simplest route to cosmological reconstruction is through
the Hubble parameter, which in a spatially flat universe is related to
the luminosity distance quite simply by \cite{st98,turner,nak99}
\beq
H(z) = \left[ \frac{d}{dz} \left( \frac{d_L(z)}{1+z} \right) \right]^{-1}.
\eeq

We may now define the dark energy density as :
\beq\label{eq:dens}
\rho_{\rm DE}=\rhoc \left[\left(\frac{H}{H_0}\right)^2-\omt (1+z)^3\right]\,\,,
\eeq
where $\rhoc=3 H^2_0/(8 \pi G)$ is the present day critical density of
an FRW universe, and $\omt$ is the present day matter density with
respect to the critical density. However, one should keep in mind a
subtle point regarding this definition of the dark energy density. Its
ambiguity lies in the value of $\omt$. From CMB and galaxy clustering
data, we obtain an estimate of the total amount of clustered
non-relativistic matter present today (denoted by $\tilde \omt$).
However, $\omt$ may be {\em different} from $\tilde \omt$ due to a
contribution from a part of the unclustered dark energy which also has
a dust-like equation of state.  Fortunately, this difference (if it
exists) appears to be small, \eg~not exceeding $0.1$ for the best-fit
shown in section (2.2).

Information extracted from SNe observations regarding $d_L(z)$
therefore translates directly into knowledge of $H(z)$, the dark
energy density, and, through \cite{saini00}
\ber\label{eq:state}
q(x) &=& - \frac{\ddot{a}}{a H^2} \equiv \frac{H^{\prime}}{H} x -1 ~, \\
w(x) &=& {2 q(x) - 1 \over 3 \left( 1 - \om(x) \right)}
\equiv \frac{(2 x /3) \ H^{\prime}/H - 1}{1 \ - \ (H_0/H)^2
\omt \ x^3} ~;~x=1+z\,\,,
\eer
into knowledge about the deceleration parameter of the universe and
the equation of state of dark energy.

For a meaningful reconstruction of DE one must construct an ansatz for
$H(z)$ which is sufficiently versatile to accommodate a large class of
DE models.  (Alternatively one could devise an ansatz for $d_L(z)$ or
$w(z)$; for a summary of different approaches see
\cite{asss03a,asss03b}.)  An ansatz which works quite well for
Quintessence and also for the Chaplygin gas and Braneworld models is
\cite{sahni03}
\beq
h(x) =  \frac{H(x)}{H_0} =
\left\lbrack \omt x^3 + A_0 + A_1x + A_2 x^2\right\rbrack^\half\,\, , \ x= 1+z\,\,,
\label{eq:taylor}
\eeq
where $A_0+A_1+A_2 = 1-\omt$. Note that this ansatz should not be
considered as a truncated Taylor series for $h^2(z)$.  Rather, it is
an interpolating fit for $h^2(z)$ having the right behaviour for small
and large values of $z$. The number of terms in this fit is sufficient
given the amount and accuracy of the present supernovae data.  With
more and better data in future, more terms with intermediate (\eg,
half-integer) powers of $x$ may be added to it. Truncated Taylor
series in powers of $z$ with any finite number of terms may not be
used to analyse most interesting quantities (including $h(z)$ and
$d_L(z)$) for the already existing supernovae data since these series
become divergent at $z\sim 1$. In particular, for the standard
$\l$CDM model ($A_1=A_2=0$) the convergence radius of the Taylor
expansions for $h(z)$ and $d_L(z)$ centered at $z=0$ is given by the
distance to the closest singular point in the complex $z$ plane which
lies at $z=(1/\omt-1)^{1/3}\exp(\pm i\pi/3) -1$.  So, for $\omt
=0.3$ the Taylor series diverges at $z\ge 1.197$.

This is equivalent to the following ansatz for DE density (with
respect to the critical density) : 
\beq
\tilde \rho_{\rm DE}(x) = \rho_{\rm DE}/\rhoc = A_0 + A_1x + A_2 x^2\,\,,
\eeq
which is exact for the cosmological constant $w = -1$ ($A_1 = A_2 =
0$) and for DE models with $w = -2/3$ ($A_0 = A_2 = 0$) and $w = -1/3$
($A_0 = A_1 = 0$). Here, we neglect a possible difference between
$\omt$ and $\tilde\omt$ mentioned above since it appears to be
unimportant for the redshifts involved ($z<2$), though it may become
important for larger $z$.

The corresponding expression for the equation of state of DE is :
\beq
w(x)=-1+\frac{A_1 x+2 A_2 x^2}{3(A_0+A_1 x+A_2 x^2)}\,\,.
\eeq

A glimpse into the properties of dark energy is also provided by a two 
parameter approximation for the equation of state :
\beq
w(z) = w_0 + w_1~z~,
\label{eq:state1}
\eeq
which can be trusted for small values of $z \lleq 1$.

\begin{figure*}
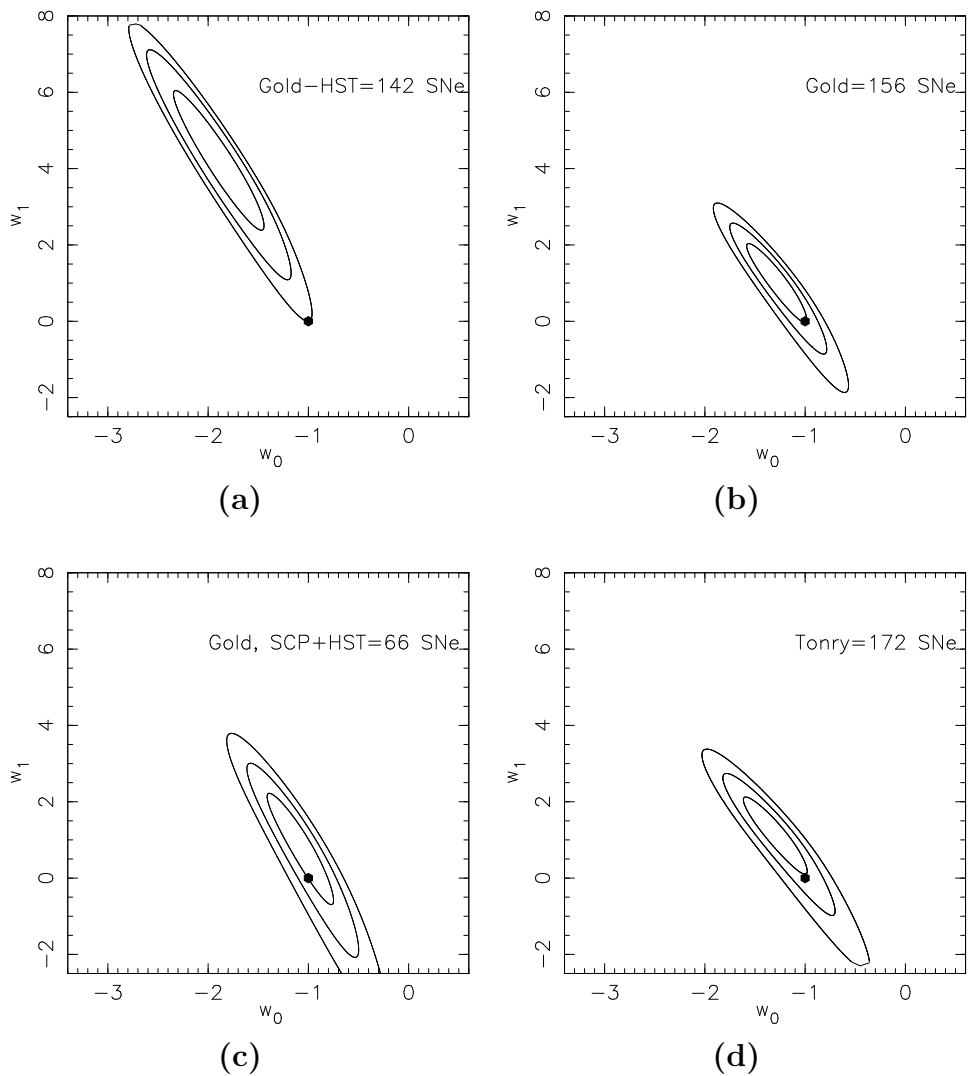
 
\centering
\begin{center}
$\begin{array}{c@{\hspace{0.2in}}c}
\multicolumn{1}{l}{\mbox{}} &
\multicolumn{1}{l}{\mbox{}} \\ 
\epsfxsize=2.4in
\epsffile{w_ell_g-hst.epsi} &  
\epsfxsize=2.4in
\epsffile{w_ell_g.epsi} \\
\mbox{\bf (a)} & \mbox{\bf (b)}
\end{array}$
$\begin{array}{c@{\hspace{0.2in}}c}
\multicolumn{1}{l}{\mbox{}} &
\multicolumn{1}{l}{\mbox{}} \\ 
\epsfxsize=2.4in
\epsffile{w_ell_scp.epsi} &  
\epsfxsize=2.4in
\epsffile{w_ell_tonry.epsi} \\  
\mbox{\bf (c)} & \mbox{\bf (d)}
\end{array}$
\end{center}
\caption{\small 
$1\sigma, \ 2\sigma, \ 3\sigma$ confidence levels in the $w_0-w_1$
space for the ansatz~(\ref{eq:state1}) for $\omt=0.3$, using different
subsets of data from \cite{riess04}. The filled circle represents the
$\l$CDM point.}
\label{fig:w_riess}
\end{figure*}

The likelihood for the parameters of the ansatz can be determined by
minimising a $\chi^2$-statistic:

\beq
\chi^2(H_0,\omt,p_j)=\sum_i \frac{[\mu_{{\rm fit},i}(z_i;H_0,\omt,p_j)-\mu_{0,i}]^2}{\sigma^2_i}\,\,,
\eeq
where $\mu_{0,i}=m_B-M=5{\rm log}d_L+25$ is the extinction corrected
distance modulus for SNe at redshift $z_i$, $\sigma_i$ is the
uncertainty in the individual distance moduli (including the
uncertainty in galaxy redshifts due to a peculiar velocity of $400$
km/s), and $p_j$ are the parameters of the relevant ansatz ($A_1, A_2$
for the ansatz (\ref{eq:taylor}) and $w_0, w_1$ for the ansatz
(\ref{eq:state1})). We assume a flat universe for our analysis but
make no further assumptions on the nature of dark energy. For most of
the following results, we marginalise over the nuisance parameter
$H_0$ by integrating the probability density ${\rm e}^{-\chi^2/2}$
over all values of $H_0$.

\begin{figure*}
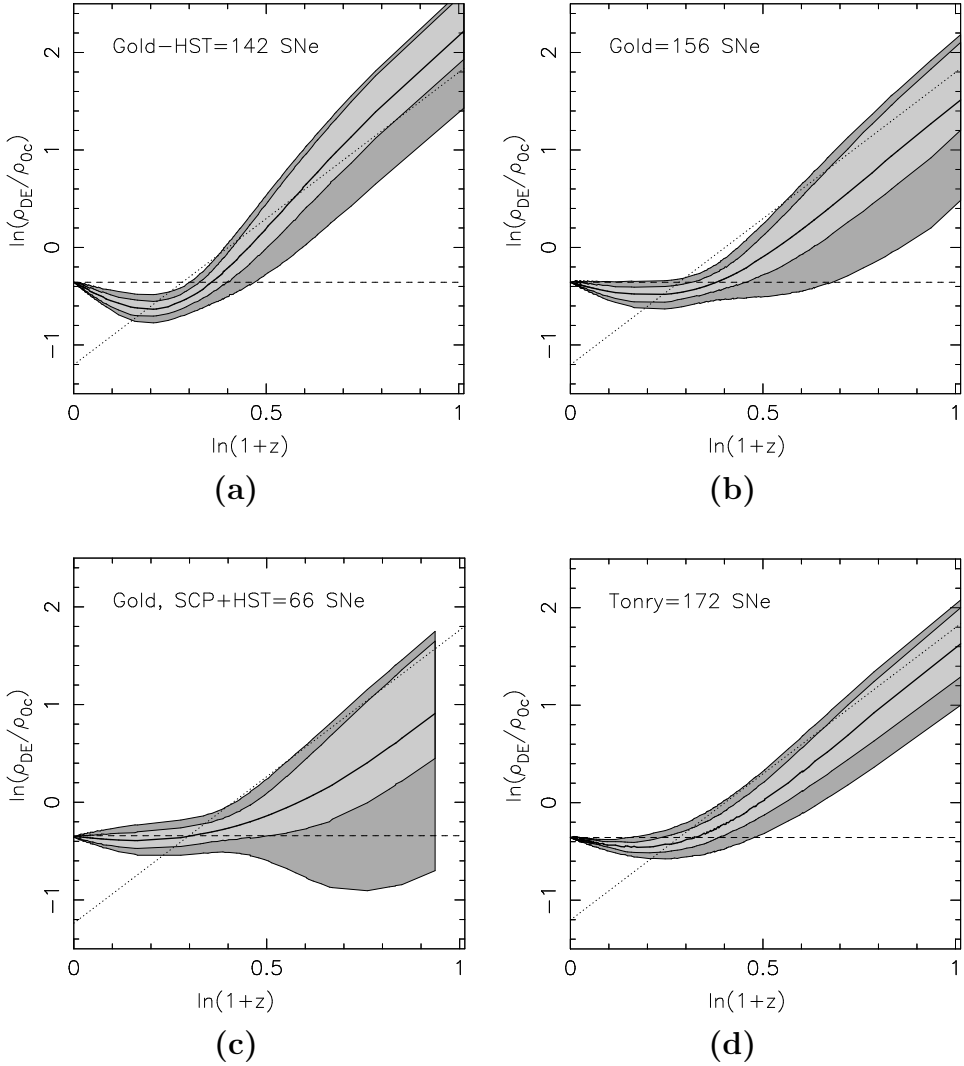
 
\centering
\begin{center}
$\begin{array}{c@{\hspace{0.2in}}c}
\multicolumn{1}{l}{\mbox{}} &
\multicolumn{1}{l}{\mbox{}} \\ 
\epsfxsize=2.4in
\epsffile{h_exp_dens_g-hst.epsi} &  
\epsfxsize=2.4in
\epsffile{h_exp_dens_g.epsi} \\
\mbox{\bf (a)} & \mbox{\bf (b)}
\end{array}$
$\begin{array}{c@{\hspace{0.2in}}c}
\multicolumn{1}{l}{\mbox{}} &
\multicolumn{1}{l}{\mbox{}} \\ 
\epsfxsize=2.4in
\epsffile{h_exp_dens_g_scp.epsi} &  
\epsfxsize=2.4in
\epsffile{h_exp_dens_tonry.epsi} \\  
\mbox{\bf (c)} & \mbox{\bf (d)}
\end{array}$
\end{center}
\caption{\small
The logarithmic variation of dark energy density $\rho_{\rm
  DE}(z)/\rhoc$ (where $\rhoc=3 H_0^2/8 \pi G$ is the present critical
energy density) with redshift for $\omt=0.3$ using different subsets
of data from \cite{riess04}, for the ansatz~(\ref{eq:taylor}). In each
panel, the thick solid line shows the best-fit, the light grey contour
represents the $1\sigma$ confidence level, and the dark grey contour
represents the $2\sigma$ confidence level around the best-fit.  The
dotted line denotes matter density $\omt (1+z)^3$, and the dashed
horizontal line denotes $\l$CDM.}
\label{fig:h_exp_dens_riess}
\end{figure*}

\begin{figure*}
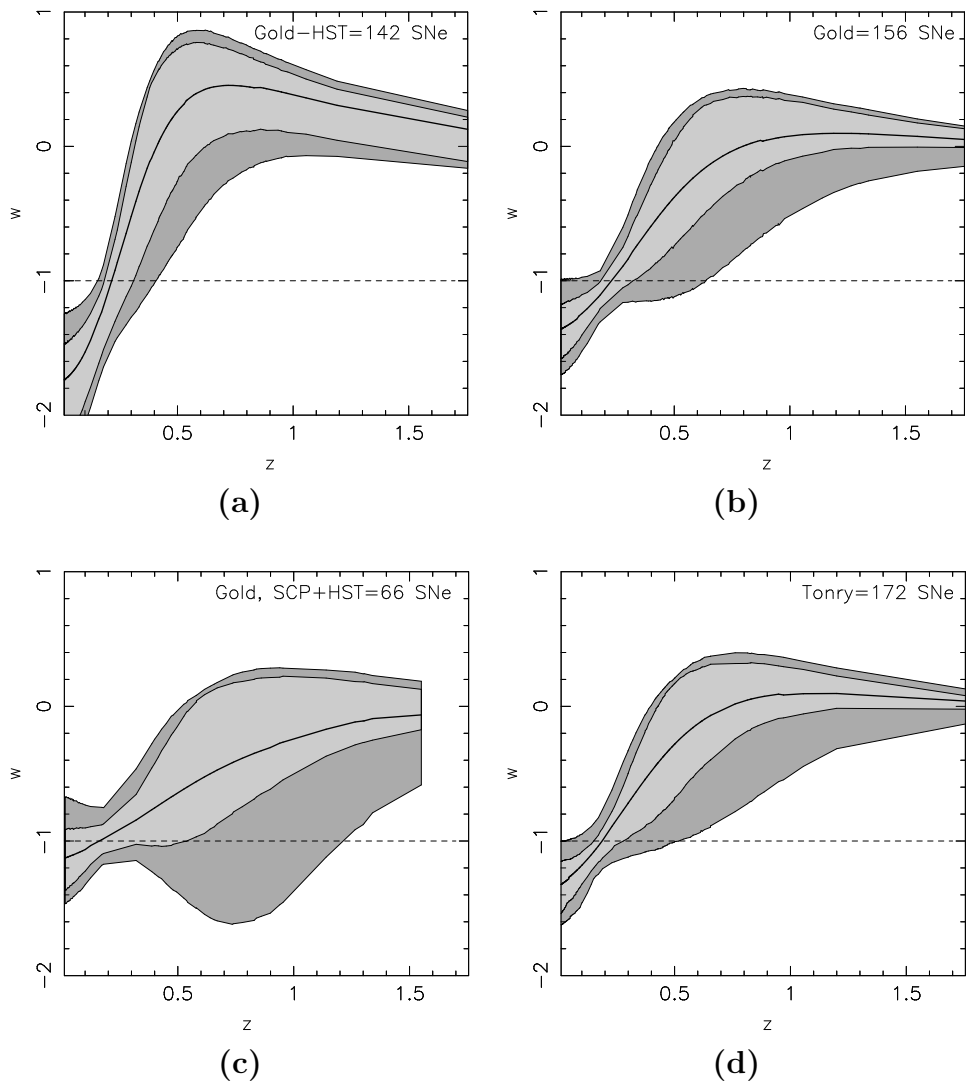
 
\centering
\begin{center}
$\begin{array}{c@{\hspace{0.2in}}c}
\multicolumn{1}{l}{\mbox{}} &
\multicolumn{1}{l}{\mbox{}} \\ 
\epsfxsize=2.4in
\epsffile{h_exp_w_g-hst.epsi} &  
\epsfxsize=2.4in
\epsffile{h_exp_w_g.epsi} \\
\mbox{\bf (a)} & \mbox{\bf (b)}
\end{array}$
$\begin{array}{c@{\hspace{0.2in}}c}
\multicolumn{1}{l}{\mbox{}} &
\multicolumn{1}{l}{\mbox{}} \\ 
\epsfxsize=2.4in
\epsffile{h_exp_w_g_scp.epsi} &  
\epsfxsize=2.4in
\epsffile{h_exp_w_tonry.epsi} \\  
\mbox{\bf (c)} & \mbox{\bf (d)}
\end{array}$
\end{center}
\caption{\small 
The variation of equation of state of dark energy $w(z)$ with redshift
for $\omt=0.3$ using different subsets of data from \cite{riess04},
for the ansatz~(\ref{eq:taylor}). In each panel, the thick solid line
shows the best-fit, the light grey contour represents the $1\sigma$
confidence level, and the dark grey contour represents the $2\sigma$
confidence level around the best-fit.  The dashed horizontal line
denotes $\l$CDM.}
\label{fig:h_exp_w_riess}
\end{figure*}

We first study the different subsamples of the new SNe data reported
in \cite{riess04} in some detail using the ansatz (\ref{eq:taylor})
and (\ref{eq:state1}). Riess \etal \cite{riess04} reanalysed the
existing SNe data compiled by different search teams (mainly the High
Redshift Search Team (HZT) and the Supernova Cosmology Project (SCP)
team) and added to these 16 new SNe observed by HST. They have
rejected many of the previously published SNe due to lack of complete
photometric record, uncertain classification, etc. They divide the
total data-set into ``high-confidence'' (`Gold') and ``likely but not
certain''(`Silver') subsets. In our calculations we will consider
their ``high-confidence'' (`Gold') subset.  Figure~\ref{fig:w_riess}
shows the $(w_0, w_1)$ confidence levels for the ansatz
(\ref{eq:state1}) using different subsets of the new data for
$\omt=0.3$.  The top two panels in fig~\ref{fig:w_riess} are similar
to the corresponding panels of figure 10 of \cite{riess04}, and the
bottom panels show other subsets of data.  The panel (a) shows the
`Gold' sample of \cite{riess04} without the 14 new HST points, and the
panel (b) shows the full `Gold' sample of 156 SNe. Of the last two
panels, the panel (c) shows the results for a subset of the `Gold'
sample, consisting only of the points obtained by the SCP team and the
new HST points, while the panel (d) shows the confidence levels for
the older data set of 172 SNe published in Tonry \etal~\cite{tonry03}.
The greatest difference is clearly between the panels (a) and (c). We
should also note that the panel (b) (comprising of the new reanalysed
data) and panel (d) (comprising of the old HZT data) are qualitatively
similar, with panel (b) having tighter errors.

We now analyse these different subsets using the
ansatz~(\ref{eq:taylor}).  In figures~\ref{fig:h_exp_dens_riess} and
~\ref{fig:h_exp_w_riess}, we show the variation of dark energy density
and dark energy equation of state obtained using the ansatz
~(\ref{eq:taylor}) for the different subsets of data, fixing
$\omt=0.3$. It is interesting to note that panels (b) and (d) in both
figures are quite similar, and we therefore conclude that the `Gold'
sample shows an evolution for DE which is consistent with that
obtained from the older sample of Tonry \etal \cite{tonry03}.

From figure~\ref{fig:w_riess} we find that for all four datasets, the
largest degeneracy direction in $w_0-w_1$ plane corresponds to the
curve $w_0 + 0.25 w_1 \simeq -1$. This immediately suggests that
$w(z=0.25) \simeq -1$, and this result appears to be quite robust
since one can also arrive at it by choosing a very different ansatz
(\ref{eq:taylor}) to determine $w(z)$, as shown in
figure~\ref{fig:h_exp_w_riess}.

As pointed out in \cite{riess04}, and confirmed by the
figures~\ref{fig:w_riess},~\ref{fig:h_exp_dens_riess},~\ref{fig:h_exp_w_riess},
the {\em maximum evolution} in DE is for the `Gold -- HST' data. Less
evolution is shown by the `Gold' data set.  However, we would like to
emphasise again that the `Gold' data set, which includes the 14 new
HST points, gives roughly the same degree of evolution for DE as the
original data reported in Tonry \etal~\cite{tonry03} and analysed in
\cite{asss03b}. Thus the results pertaining to the evolution of DE
reported in Paper I remain valid also for the new SNe (Gold) data set.

Note that even the maximum rate of evolution of DE obtained in
\cite{asss03b} in terms of $w$ is ${dw\over dz}\sim $ {\it few} which
is in complete agreement with the results obtained in \cite{riess04}.
Does this mean however that the new supernovae data completely exclude
models with a fast phase transition in dark energy
\cite{bassett,corasaniti} for which $|{dw\over dz}|\gg 1$ over a
narrow range of redshift $\delta z \ll 1$?  In our view such a
conclusion would be premature.  The reason is that using an ansatz
such as (\ref{eq:taylor}) or (\ref{eq:state1}) (as in \cite{riess04})
is equivalent to smoothing the evolution of the Universe over a
redshift interval $\Delta z \propto 1/N$, $N$ being the number of free
parameters used in the ansatz.  Clearly, after such an implicit
smoothing fast phase transitions in dark energy will disappear not so
much because they disagree with the data but because of the manner in
which the data has been analysed (see \cite{asss03b} for a more
detailed discussion of these issues).

Figures~\ref{fig:w_riess},~\ref{fig:h_exp_dens_riess},
and~\ref{fig:h_exp_w_riess} also illustrate that the degree of DE
evolution can be quite sensitive to the manner in which the SNe data
is sampled.  Comparing panels (a) and (c) in these figures, we find
that the degree of evolution of DE is largest for the `Gold -- HST'
data set and least for the `Gold, SCP + HST' data.  (The latter is in
better agreement with $\l$CDM than the other three data sets.)

\subsection{Analysis of `Gold' SNe dataset :}

\begin{figure*}
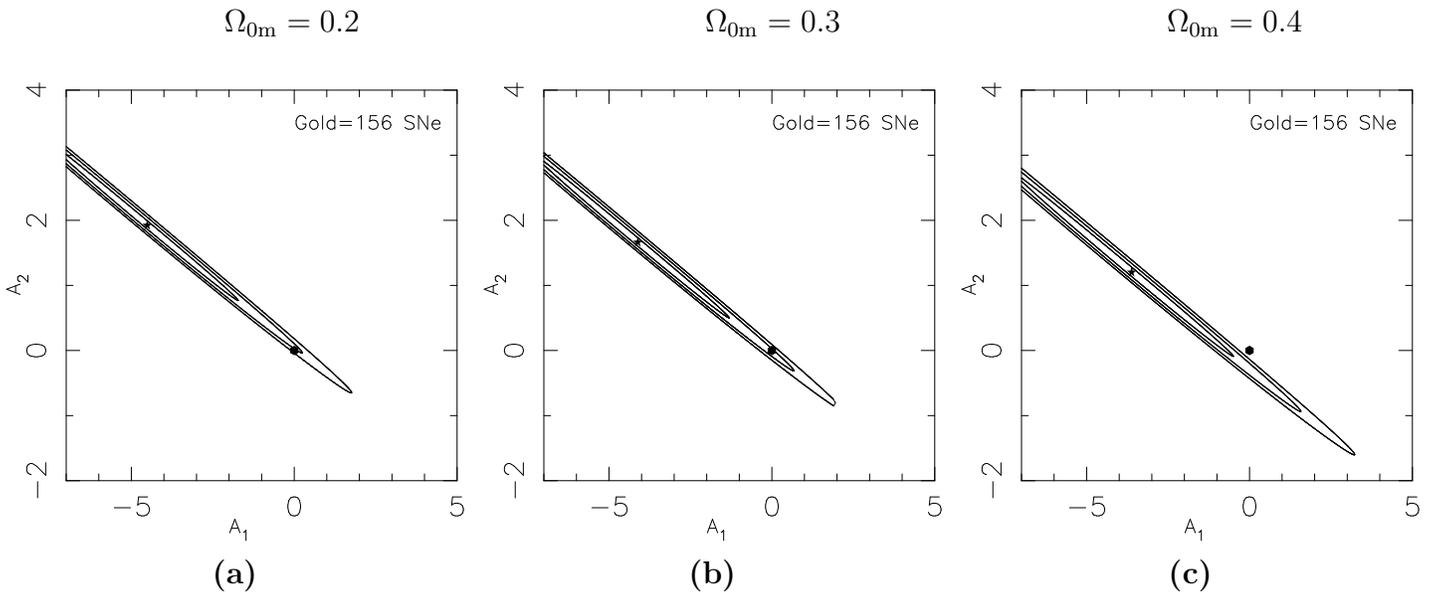

\centering
\begin{center}
\vspace{0.05in}
\centerline{\mbox{\hspace{0.7in} $\omt=0.2$ \hspace{1.7in} $\omt=0.3$ \hspace{1.6in} $\omt=0.4$}}
$\begin{array}{@{\hspace{-0.5in}}c@{\hspace{0.1in}}c@{\hspace{0.1in}}c}
\multicolumn{1}{l}{\mbox{}} &
\multicolumn{1}{l}{\mbox{}} &
\multicolumn{1}{l}{\mbox{}} \\ [-0.5cm]
\epsfxsize=2.4in
\epsffile{h_exp_ell_g_mi.epsi} &
\epsfxsize=2.4in
\epsffile{h_exp_ell_g.epsi} &
\epsfxsize=2.4in
\epsffile{h_exp_ell_g_pl.epsi} \\
\mbox{\bf (a)} & \mbox{\bf (b)} & \mbox{\bf (c)}
\end{array}$
\end{center}
\caption{\small
The ($A_1,A_2$) parameter space for the ansatz (\ref{eq:taylor}) for
different values of $\omt$, using the `Gold' sample of SNe from
\cite{riess04}. The star in each panel marks the best-fit point, and
the solid contours around it mark the $1\sigma, 2\sigma, 3\sigma$
confidence levels around it. The filled circle represents the
$\l$CDM point. The corresponding $\chi^2$ for the best-fit points
are given in table~\ref{tab:chi}. }
\label{fig:h_exp_ell}
\end{figure*}

We now examine the `Gold' data set in some detail using the polynomial
expansion of dark energy, (\ref{eq:taylor}). In the
figure~\ref{fig:h_exp_ell} we show the confidence levels for the
parameters $(A_1,A_2)$ of the ansatz for three different values of
$\omt$. The $\chi^2$ value for the best-fit in each case is given in
table~\ref{tab:chi}. The $\chi^2$ values for the corresponding
$\l$CDM models are also given for comparison. Interestingly, in
all three cases the confidence ellipse has the same inclination, it
only appears to shift downwards as $\omt$ increases.

\begin{table}
\begin{center}
\caption{
$\chi^2$ per degree of freedom for best-fit and $\l$CDM models
for analysis using the `Gold' sample of SNe from \cite{riess04}.
$w_0$ is the present value of the equation of state of dark energy in
best-fit models. } \vspace{0.2cm}
\label{tab:chi}
\begin{tabular}{cccc}
$ $&\multicolumn{2}{c}{Best-fit}&$\l$CDM \\
$\omt$&$ w_0$&$\chi^2_{\rm min}$&$ \chi^2$\\
\hline
$0.20$&$-1.20$&$1.036$&$1.109$\\
$0.30$&$-1.35$&$1.034$&$1.053$\\
$0.40$&$-1.59$&$1.030$&$1.086$\\
\end{tabular}
\end{center}
\end{table}

\begin{figure*}
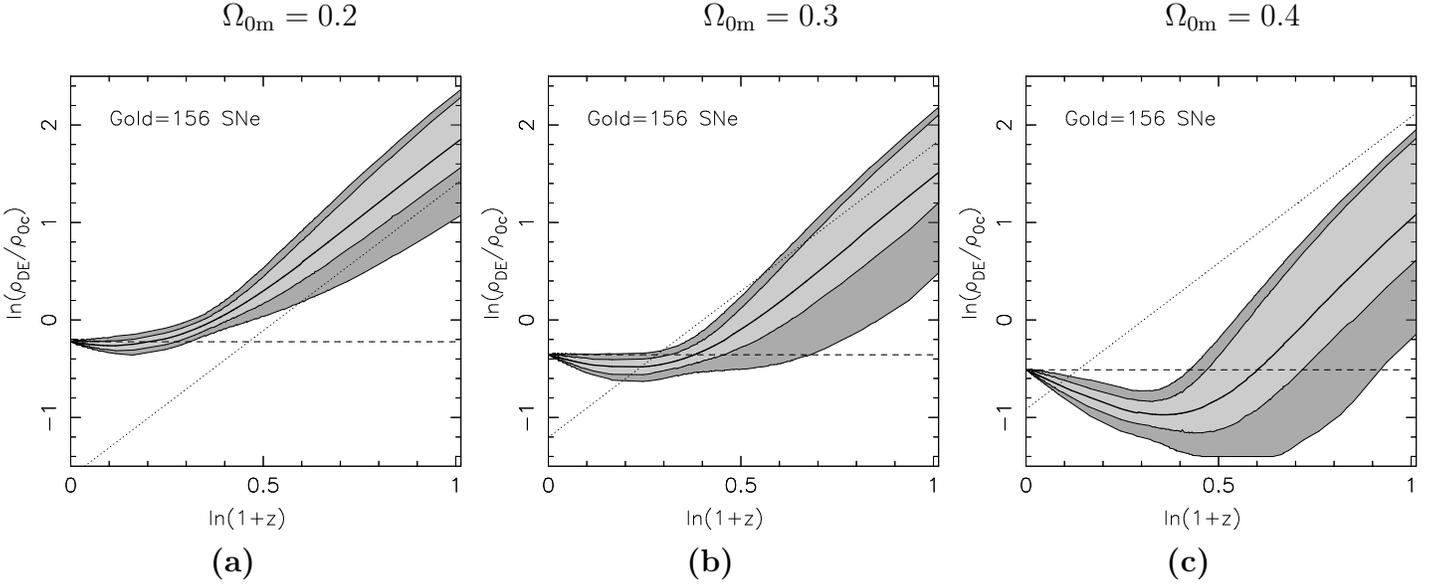

\centering
\begin{center}
\vspace{0.05in}
\centerline{\mbox{\hspace{0.7in} $\omt=0.2$ \hspace{1.7in} $\omt=0.3$ \hspace{1.6in} $\omt=0.4$}}
$\begin{array}{@{\hspace{-0.5in}}c@{\hspace{0.1in}}c@{\hspace{0.1in}}c}
\multicolumn{1}{l}{\mbox{}} &
\multicolumn{1}{l}{\mbox{}} &
\multicolumn{1}{l}{\mbox{}} \\ [-0.5cm]
\epsfxsize=2.4in
\epsffile{h_exp_dens_g_mi.epsi} &
\epsfxsize=2.4in
\epsffile{h_exp_dens_g.epsi} &
\epsfxsize=2.4in
\epsffile{h_exp_dens_g_pl.epsi} \\
\mbox{\bf (a)} & \mbox{\bf (b)} & \mbox{\bf (c)}
\end{array}$
\end{center}
\caption{\small
The logarithmic variation of dark energy density $\rho_{\rm
  DE}(z)/\rhoc$ (where $\rhoc=3 H_0^2/8 \pi G$ is the present critical
energy density) with redshift for different values of $\omt$, using
the `Gold' sample of SNe from \cite{riess04}. The reconstruction is
done using the polynomial fit to dark energy,
ansatz~(\ref{eq:taylor}). In each panel, the thick solid line shows
the best-fit, the light grey contour represents the $1\sigma$
confidence level, and the dark grey contour represents the $2\sigma$
confidence level around the best-fit.  The dotted line denotes matter
density $\omt (1+z)^3$, and the dashed horizontal line denotes
$\l$CDM.}
\label{fig:h_exp_dens}
\end{figure*}

\begin{figure*}
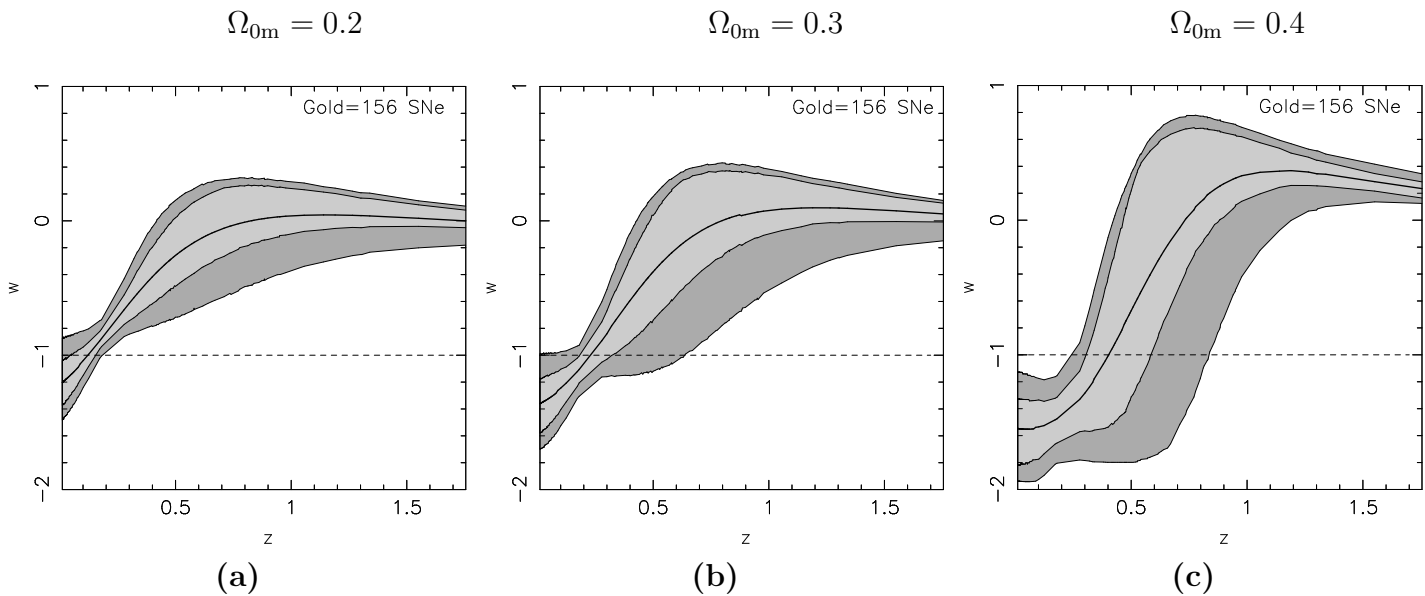

\centering
\begin{center}
\vspace{0.05in}
\centerline{\mbox{\hspace{0.7in} $\omt=0.2$ \hspace{1.7in} $\omt=0.3$ \hspace{1.6in} $\omt=0.4$}}
$\begin{array}{@{\hspace{-0.5in}}c@{\hspace{0.1in}}c@{\hspace{0.1in}}c}
\multicolumn{1}{l}{\mbox{}} &
\multicolumn{1}{l}{\mbox{}} &
\multicolumn{1}{l}{\mbox{}} \\ [-0.5cm]
\epsfxsize=2.4in
\epsffile{h_exp_w_g_mi.epsi} &
\epsfxsize=2.4in
\epsffile{h_exp_w_g.epsi} &
\epsfxsize=2.4in
\epsffile{h_exp_w_g_pl.epsi} \\
\mbox{\bf (a)} & \mbox{\bf (b)} & \mbox{\bf (c)}
\end{array}$
\end{center}
\caption{\small
The evolution of $w(z)$ with redshift for different values of $\omt$,
for the `Gold' sample of SNe from \cite{riess04}. The reconstruction
is done using the polynomial fit to dark energy,
equation~(\ref{eq:taylor}). In each panel, the thick solid line shows
the best-fit, the light grey contour represents the $1\sigma$
confidence level, and the dark grey contour represents the $2\sigma$
confidence level around the best-fit. The dashed line represents
$\l$CDM.  }
\label{fig:h_exp_w}
\end{figure*}

In figure~\ref{fig:h_exp_dens}, we show the variation of the dark
energy density with redshift for different values of the current
matter density. We see that, for higher $\omt$, the dark energy density
evolution is sharper. The reader should also note that the growth of
$\tilde \rho_{\rm DE}$ with time in the panels (b) and (c) is indicative of the
phantom nature of DE ($w \leq -1$) at recent times ($z \lleq 0.25$ for
$\omt = 0.3$ and $z \lleq 0.4$ for $\omt = 0.4$, see
figure~\ref{fig:h_exp_w}).

We may obtain more information from the dark energy density by
considering a weighted average of the equation of state:
\beq
1+\bar{w} = \frac{1}{\Delta \ {\rm ln}(1+z)} \int [1+w(z)] d \ {\rm ln}(1+z) = \frac{1}{3} \frac{\Delta \ {\rm ln}\tilde \rho_{\rm DE}}{\Delta \ {\rm ln}(1+z)}  \,\,,
\label{eq:w_avg}
\eeq
where $\Delta$ denotes the total change of the variable between
integration limits.  Thus the variation in the dark energy density
depicted in figure~\ref{fig:h_exp_dens} is very simply related to the
weighted average equation of state of dark energy. The value of
$\bar{w}$ for different ranges of integration are shown in
table~\ref{tab:w_avg}. We have taken the ranges of integration to be
approximately equally spaced in ${\rm ln}(1+z)$. In all three cases
shown, the value of $\bar{w}$ changes noticeably from close to $-1$ in
the first bin to close to zero in the second bin. This indicates that
the equation of state of DE is evolving from $w \lleq -1$ today to $w
\simeq 0$ at $z \simeq 1$. Note that these results are in very good
agreement with those reported in Table 1 of Paper I.

\begin{table}
\begin{center}
\caption{
The   weighted  average $\bar{w}$  (eq~\ref{eq:w_avg}) over  specified
redshift  ranges for  analysis using  the  `Gold'  sample of SNe  from
\cite{riess04}. The best-fit value and  $1\sigma$ deviations from  the
best-fit are shown.}  \vspace{0.2cm}
\label{tab:w_avg}
\begin{tabular}{cccc}
&\multicolumn{3}{c}{$\bar{w}$}\\
$\omt$&$\Delta z=0-0.414$&$\Delta z=0.414-1$&$\Delta z=1-1.755$\\
\hline
$0.2$&$-0.847^{+0.019}_{-0.043}$&$-0.118^{+0.280}_{-0.211}$&$0.089^{+0.067}_{-0.039}$\\
$0.3$&$-1.053^{+0.089}_{-0.070}$&$-0.159^{+0.319}_{-0.259}$&$0.118^{+0.073}_{-0.041}$\\
$0.4$&$-1.310^{+0.220}_{-0.179}$&$-0.210^{+0.452}_{-0.340}$&$0.215^{+0.081}_{-0.050}$\\
\end{tabular}
\end{center}
\end{table}

Figure~\ref{fig:h_exp_w} shows the corresponding variation of the DE
equation of state with redshift for different $\omt$. Here also, there
is strong evidence for evolution of DE.  We see that for higher values
of $\omt$, the dark energy equation of state has a more negative value
at present and shows a sharper evolution over redshift. It is
interesting to note that a constant equation of state which is $<-1$
during the evolution of the universe from $z \simeq 1.7$ to $z=0$
appears not to be supported by the recent SNe data.

To summarize, our results clearly demonstrate that evolving DE is by
no means excluded by the most recent SNe observations. On the
contrary, our results for $\tilde \rho_{\rm DE}$ and $w_{\rm DE}$ obtained
using the `Gold' sample of \cite{riess04}
(figures~\ref{fig:h_exp_dens},~\ref{fig:h_exp_w}) are very similar to
the results which we obtain using the SNe samples of
\cite{tonry03,barris03}.  For $0.2 \leq \omt \leq 0.4$ the best-fit DE
model evolves from $w \lleq -1$ at $z \simeq 0$ to $w \simeq 0$ at $z
\simeq 1$ in agreement with the results of Paper I.

\subsection{DE reconstruction using SNe(`Gold')+CMB :}

\begin{figure*}
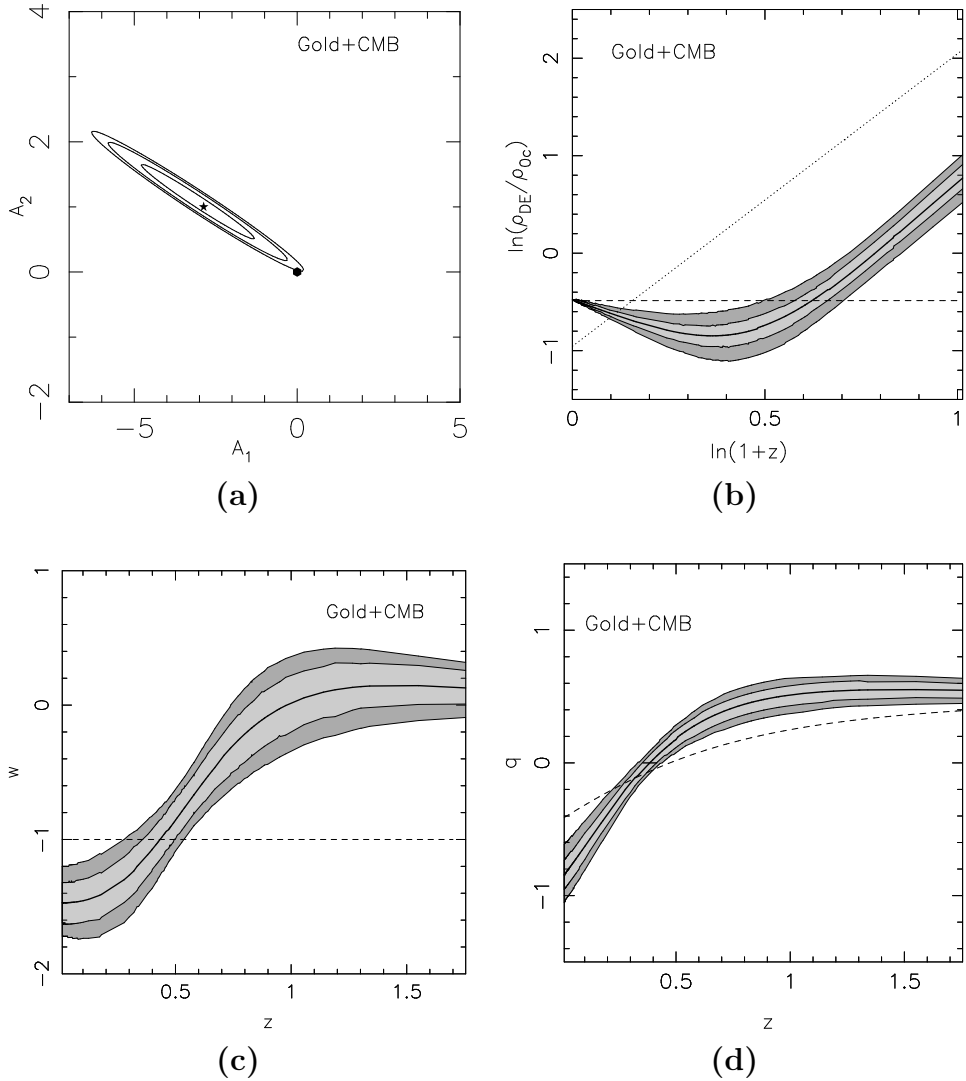
 
\centering
\begin{center}
$\begin{array}{c@{\hspace{0.2in}}c}
\multicolumn{1}{l}{\mbox{}} &
\multicolumn{1}{l}{\mbox{}} \\ 
\epsfxsize=2.4in
\epsffile{h_exp_ell_g_cmb.epsi} &  
\epsfxsize=2.4in
\epsffile{h_exp_dens_g_cmb.epsi} \\
\mbox{\bf (a)} & \mbox{\bf (b)}
\end{array}$
$\begin{array}{c@{\hspace{0.2in}}c}
\multicolumn{1}{l}{\mbox{}} &
\multicolumn{1}{l}{\mbox{}} \\ 
\epsfxsize=2.4in
\epsffile{h_exp_w_g_cmb.epsi} &  
\epsfxsize=2.4in
\epsffile{h_exp_dec_g_cmb.epsi} \\  
\mbox{\bf (c)} & \mbox{\bf (d)}
\end{array}$
\end{center}
\caption{\small 
Results from analysis of SNe(Gold)+CMB data, using
ansatz~(\ref{eq:taylor}). $\omt$ and $h$ are fixed at best-fit values
of $\omt = 0.385, h = 0.60$. Panel (a) shows the confidence levels in
the $(A_1,A_2)$ parameter space. The star marks the best-fit and the
filled circle marks the $\l$CDM point. Panel (b) shows the logarithmic
variation of dark energy density with redshift. Panel (c) shows the
evolution of dark energy equation of state with redshift. Panel (d)
shows the variation of the deceleration parameter with redshift. In
all three panels (b), (c) and (d), the thick solid line represents the
best-fit, the light grey contours represent the $1\sigma$ confidence
level, and the dark grey contours represent the $2\sigma$ confidence
levels. The dashed line in panels (b), (c) and (d) represents $\l$CDM,
and the dotted line in panel (b) represents the matter density. The
horizontal thick solid line in (d) represents the $1\sigma$ limits on
the transition redshift at which the universe starts accelerating.}
\label{fig:h_exp_cmb}
\end{figure*}

Observations of the cosmic microwave background and Type Ia supernovae
provide us with complementary insight into the nature of dark energy
\cite{bond,doran,bean02,hu01}.  We may use the WMAP result of $R =
\sqrt{\omt}\int_0^{z_{\rm ls}}dz/h(z) = 1.710 \pm 0.137$ (from WMAP
data alone) in conjunction with the SNe `Gold' sample to reconstruct
DE. For this purpose, we use $\Omega_b h^2=0.024$ and $\omt h^2 =0.14
\pm 0.02$ \cite{spergel03}. To calculate $z_{ls}$ we use a fitting
function given in \cite{hu}:
\beq
z_{ls}=1048 [1+0.00124 (\Omega_b h^2)^{-0.738}][1+g_1 (\omt h^2)^{g_2}]\,\,,
\eeq
where 
\ber
g_1 &=& 0.078(\Omega_b h^2)^{-0.238} [1+39.5 (\Omega_b h^2)^{0.763}]^{-1},\\
g_2 &=& 0.56 [1+21.1 (\Omega_b h^2)^{1.81}]^{-1}\,\,.
\eer

\begin{table}
\begin{center}
\caption{
The weighted average $\bar{w}$ (eq~\ref{eq:w_avg}) over specified
redshift ranges for analysis from SNe+CMB data. The best-fit value and
$1\sigma$ deviations from the best-fit are shown.}  \vspace{0.2cm}
\label{tab:w_cmb}
\begin{tabular}{cccc}
&\multicolumn{3}{c}{$\bar{w}$}\\
$\omt$&$\Delta z=0-0.414$&$\Delta z=0.414-1$&$\Delta z=1-1.755$\\
\hline
$0.385$&$-1.287^{+0.016}_{-0.056}$&$-0.229^{+0.070}_{-0.177}$&$0.142^{+0.051}_{-0.033}$\\
\end{tabular}
\end{center}
\end{table}

Figure~\ref{fig:h_exp_cmb} shows our reconstruction of DE obtained
with the ansatz~(\ref{eq:taylor}) using the WMAP result together with
the SNe `Gold' sample. The best fit values for this reconstruction
are: $\omt = 0.385, A_1 =-2.87 , A_2 = 1.01, h = 0.60$. The best fit
dark energy density is $\tilde \rho_{\rm DE}(x)=2.475-2.87 x+1.01
x^2$. Note that the best fit $\tilde \rho_{\rm DE}$ decreases
monotonically with redshift up to $z \simeq 0.4$ and then begins to
increase. This reflects the phantom nature of DE ($w < -1$) at lower
redshifts. The equation of state behaves as before, evolving from $w_0
\lleq -1$ to $w \simeq 0$ at $z \simeq 1$. From the
figure~\ref{fig:h_exp_cmb}(d), we see that the deceleration parameter
$q$ has a value of $q_0 = -0.84 \pm 0.11$ at present. The transition
from deceleration to acceleration ($q(z_T)=0$)occurs at a redshift of
$z_T = 0.39 \pm 0.03$. Therefore, from a joint analysis of CMB and SNe
data, one may obtain a fairly good idea of when the universe began to
accelerate. These results demonstrate that the best fit to SNe+CMB
observations favours evolving DE with a somewhat higher value of
$\omt$ and a slightly lower value of $h$. Note that the value of
$\omt$ is larger than the WMAP result $\omt = 0.27 \pm 0.04$ obtained
using the $\l$CDM prior, but is in agreement with the results obtained
in \cite{cluster1,cluster2} ($\omt \simeq 0.35 \pm 0.12$) using high
redshift clusters. Also note that, because of the dust-like behaviour
of DE at higher redshifts, the above value of $\omt$ obtained using
equations~(\ref{eq:dens}-\ref{eq:taylor}) maybe somewhat larger than
the values obtained for $\omt$ from clustering measurements.  The
value of $h \simeq 0.60$ obtained for the best fit is in tension with
the $\l$CDM based result from WMAP, $h = 0.73 \pm 0.03$, but is in
agreement with the observations of
\cite{saha,tammann,shanks,freedman,battistelli,reese,mason} which can
accommodate lower values of $h \sim 0.6$ (see also the discussion in
\cite{silk04} in this context). We therefore conclude that a joint
analysis of SNe and CMB data favours evolving DE over $\l$CDM if no
priors are placed on $\omt$ and $h$ separately ($\omt h^2 =0.14 \pm
0.02$ is assumed).

\begin{figure*}
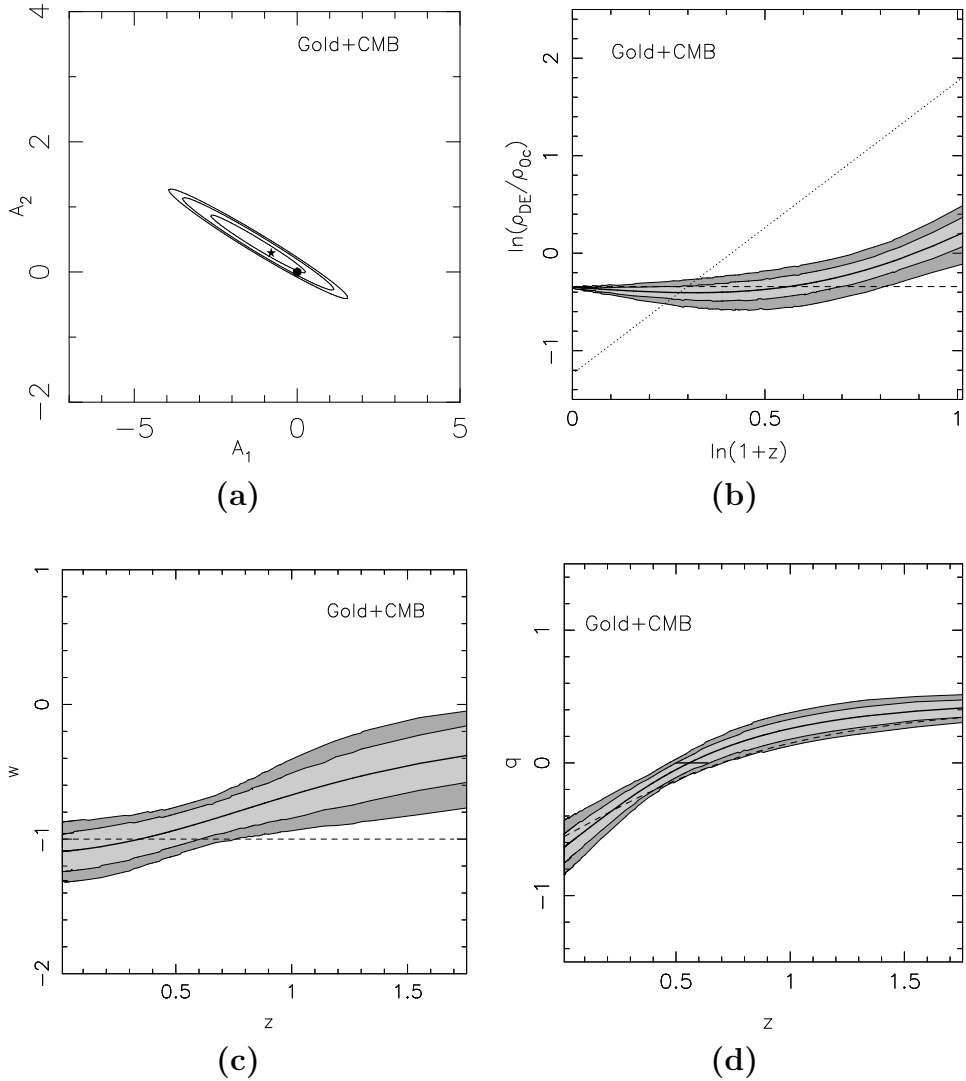
 
\centering
\begin{center}
$\begin{array}{c@{\hspace{0.2in}}c}
\multicolumn{1}{l}{\mbox{}} &
\multicolumn{1}{l}{\mbox{}} \\ 
\epsfxsize=2.4in
\epsffile{h_exp_ell_g_cmb_constr.epsi} &  
\epsfxsize=2.4in
\epsffile{h_exp_dens_g_cmb_constr.epsi} \\
\mbox{\bf (a)} & \mbox{\bf (b)}
\end{array}$
$\begin{array}{c@{\hspace{0.2in}}c}
\multicolumn{1}{l}{\mbox{}} &
\multicolumn{1}{l}{\mbox{}} \\ 
\epsfxsize=2.4in
\epsffile{h_exp_w_g_cmb_constr.epsi} &  
\epsfxsize=2.4in
\epsffile{h_exp_dec_g_cmb_constr.epsi} \\  
\mbox{\bf (c)} & \mbox{\bf (d)}
\end{array}$
\end{center}
\caption{\small 
Results for analysis of SNe(Gold)+CMB data with $\l$CDM-based WMAP
priors of $\omt=0.27 \pm 0.04$ and $h=0.71 \pm 0.06$, using
ansatz~(\ref{eq:taylor}). Panel (a) shows the confidence levels in the
$(A_1,A_2)$ space. The star marks the best-fit and the filled circle
marks the $\l$CDM point. Panel (b) shows the logarithmic variation of
dark energy density with redshift. Panel (c) shows the evolution of
dark energy equation of state with redshift.  Panel (d) shows the
variation of the deceleration parameter with redshift. In all three
panels (b), (c) and (d), the thick solid line represents the best-fit,
the light grey contours represent the $1\sigma$ confidence level, and
the dark grey contours represent the $2\sigma$ confidence levels. The
dashed line in panels (b), (c) and (d) represents $\l$CDM, and the
dotted line in panel (b) represents the matter density. The horizontal
thick solid line in (d) represents the $1\sigma$ limits on the
transition redshift at which the universe starts accelerating.}
\label{fig:h_exp_cmb_constr}
\end{figure*}

It is however a useful exercise to see how the behaviour of DE would
change if strong priors were imposed on $\omt$ and $h$. We therefore
show results obtained by using the $\l$CDM based CMB priors
\cite{spergel03}: $\omt = 0.27 \pm 0.04$ and $h = 0.71 \pm 0.06$. The
best fit in this case has $\omt=0.29$ and a dark energy density of
$\tilde \rho_{\rm DE}=1.23-0.81 x+0.29 x^2$. The equation of state at
present is $w_0 = -1.10$, and it slowly evolves to $w(z=1.75)\simeq
-0.4$. The deceleration parameter has a value of $q_0 = -0.63 \pm
0.12$ at present and the redshift at which the universe begins to
accelerate is $z_T = 0.57 \pm 0.07$. Figure~\ref{fig:h_exp_cmb_constr}
demonstrates that the time evolution of DE is {\it extremely weak} in
this case and is in good agreement with $\l$CDM cosmology (see also
\cite{wangteg}).

\section{Conclusions}

In conclusion, we find that the case for evolving dark energy
(originally demonstrated in Paper I) is upheld by the new supernova
data if no priors are imposed on $\omt$ and $h$. For a reasonable
range of $0.2 \leq \omt \leq 0.4$, the equation of state of dark
energy evolves from $w_0 < -1$ today to $w_0 \simeq 0$ at $z \simeq
1$. The above result remains in place if we add CMB priors to the
analysis. In this case, evolving dark energy with $\omt \simeq 0.385$
and $h \simeq 0.6$ is favoured over $\l$CDM, and the epoch at which
the universe began to accelerate is $z_T = 0.39 \pm 0.03$ within
$1\sigma$.  However, if we assume strong priors on $\omt$ and $h$
using the $\l$CDM based WMAP results, then the best-fit chooses an
$\omt = 0.29$. The evolution in the equation of state becomes weaker
and is in much better agreement with $\l$CDM. The redshift of
transition from deceleration to acceleration is $z_T = 0.57 \pm 0.07$,
which is closer to the $\l$CDM value of $z_T \simeq 0.7$ for this
value of $\omt$. Finally, one must note that the DE evolution becomes
weaker or stronger depending on the subsampling of the SNe dataset. A
larger number of supernovae at high redshifts, as well as better
knowledge of the values of $H_0$ and $\omt$ are therefore required
before firm conclusions are drawn about the nature of dark energy.

\section{Acknowledgements}

We would like to thank Arman Shafieloo for useful discussions. UA
thanks the CSIR for providing support for this work.  AS was partially
supported by the Russian Foundation for Basic Research, grant
02-02-16817, and by the Research Program ``Astronomy'' of the Russian
Academy of Sciences.

\end{document}